\documentclass[11pt,a4paper]{article}
\pdfoutput=1
\usepackage{jheppub}
\usepackage[english]{babel}
\usepackage[T1]{fontenc}
\usepackage[ansinew]{inputenc}
\usepackage{color}
\usepackage{graphicx}
\usepackage{amsmath,amssymb,mathrsfs}
\usepackage{fancyhdr}
\usepackage{array,bbm}
\usepackage{booktabs}
\usepackage{epstopdf}
\usepackage[font=normalsize]{subfig}
\usepackage{multirow}
\usepackage{enumerate}
\usepackage{feynmp}

\newcommand*{\trans}{\mathrm{T}}                     
\newcommand*{\unitmatrix}{\mathbbm{1}}
\newcommand*{\tvec}[1]{\ensuremath{\boldsymbol{\mathrm{#1}}}}           

\makeatletter       
\renewcommand{\thesection}{\arabic{section}}
\renewcommand{\p@subsection}{}

\makeatother

\DeclareMathOperator{\diag}{diag}

\begin{document}


\preprint{PREPRINT}

\title{Towards testing a two-Higgs-doublet model with maximal CP symmetry at the LHC:\\ construction of a Monte Carlo event generator}

%

\author[a]{J. Brehmer,}
\author[a]{V. Lendermann,}
\author[b]{M. Maniatis,}
\author[c]{O. Nachtmann,}
\author[a]{\\H.-C. Schultz-Coulon,}
\author[a]{and R. Stamen}
\affiliation[a]{
	Kirchhoff-Institut f\"ur Physik, 
	Universit\"at~Heidelberg,\\
	Im Neuenheimer Feld 226, 69120~Heidelberg, 
	Germany
}
\affiliation[b]{
	Fakult\"at f\"ur Physik,
	Universit\"at Bielefeld,\\
	Universit\"atsstrasse, 33615~Bielefeld,
	Germany
}
\affiliation[c]{
	Institut f\"ur Theoretische Physik, 
	Universit\"at~Heidelberg,\\
	Philosophenweg~16, 69120~Heidelberg, 
	Germany
}
\emailAdd{johann.brehmer@kip.uni-heidelberg.de}
\emailAdd{victor@kip.uni-heidelberg.de}
\emailAdd{maniatis@physik.uni-bielefeld.de}
\emailAdd{o.nachtmann@thphys.uni-heidelberg.de}
\emailAdd{coulon@kip.uni-heidelberg.de}
\emailAdd{stamen@kip.uni-heidelberg.de}

\abstract{
	A Monte Carlo event generator is constructed for a two-Higgs-doublet model with maximal CP symmetry, the MCPM. The model contains five physical Higgs bosons; the $\rho'$, behaving similarly to the standard-model Higgs boson, two extra neutral bosons $h'$ and $h''$, and a charged pair $H^\pm$. The special feature of the MCPM is that, concerning the Yukawa couplings, the bosons $h'$, $h''$ and $H^\pm$ couple directly only to the second generation fermions but with strengths given by the third-generation-fermion masses. Our event generator allows the simulation of the Drell-Yan-type production processes of $h'$, $h''$ and $H^\pm$ in proton-proton collisions at LHC energies. Also the subsequent leptonic decays of these bosons into the $\mu^+ \mu^-$, $\mu^+ \nu_\mu$ and $\mu^- \bar \nu_\mu$ channels are studied as well as the dominant background processes. We estimate the integrated luminosities needed in $p p$ collisions at center-of-mass energies of 8~TeV and 14~TeV for significant observations of the Higgs bosons $h'$, $h''$ and $H^\pm$ in these muonic channels.
}


\maketitle


\section{Introduction}

The experimental investigation of the Higgs sector of particle physics is one of the main aims of the LHC experiments. In the standard model (SM) there is only one physical Higgs boson. But more complicated Higgs sectors are by no means excluded. In this article we present a Monte Carlo event generator for a particular two-Higgs-doublet model. The construction of such an event generator is necessary for allowing realistic comparisons of theory and experiment to be done for this model.

Our paper is organised as follows. In section 2 we recall the main construction principles and properties of the "maximally CP-symmetric model" (MCPM) which will be studied. In section 3 we present predictions for the LHC experiments. In section 4 the implementation of the MCPM into the Monte Carlo event-generation package MadGraph and its validation are summarized. Section 5 deals with a brief analysis of MCPM signatures using the new Monte Carlo tools. We draw our conclusions in section 6.


\section{The Maximally CP-Symmetric Model}

Extending the Standard Model Higgs sector to two Higgs doublets,
\begin{equation}
\varphi_1=
\begin{pmatrix} \varphi^+_1\\ \varphi^0_1 \end{pmatrix},
\qquad
\varphi_2=
\begin{pmatrix} \varphi^+_2\\ \varphi^0_2 \end{pmatrix} \,,
\end{equation}
gives the two-Higgs-doublet model~(THDM).
There, the potential may contain many
more terms than in the SM; see e.g.~\cite{Gunion:1989we,Gunion:1992hs}.
The most general THDM Higgs potential 
can be written as follows~\cite{Haber:1993an}
\begin{multline}
\label{V_fields}
V =
m_{11}^2 (\varphi_1^\dagger \varphi_1) +
m_{22}^2 (\varphi_2^\dagger \varphi_2) -
m_{12}^2 (\varphi_1^\dagger \varphi_2) -
(m_{12}^2)^* (\varphi_2^\dagger \varphi_1)
\\
+\frac{1}{2} \lambda_1 (\varphi_1^\dagger \varphi_1)^2
+ \frac{1}{2} \lambda_2 (\varphi_2^\dagger \varphi_2)^2
+ \lambda_3 (\varphi_1^\dagger \varphi_1)(\varphi_2^\dagger \varphi_2)
\\
+ \lambda_4 (\varphi_1^\dagger \varphi_2)(\varphi_2^\dagger \varphi_1)
+ \frac{1}{2} [\lambda_5 (\varphi_1^\dagger \varphi_2)^2 + \lambda_5^*
(\varphi_2^\dagger \varphi_1)^2]
\\
+ [\lambda_6 (\varphi_1^\dagger \varphi_2) + \lambda_6^*
(\varphi_2^\dagger \varphi_1)] (\varphi_1^\dagger \varphi_1) + [\lambda_7 (\varphi_1^\dagger
\varphi_2) + \lambda_7^* (\varphi_2^\dagger \varphi_1)] (\varphi_2^\dagger \varphi_2)\;,
\end{multline}
with $m_{11}^2$, $m_{22}^2$, $\lambda_{1,2,3,4}$ real and
$m_{12}^2$, $\lambda_{5,6,7}$ complex.
Many properties of THDMs turn out 
to have a simple geometric meaning if we introduce
gauge invariant
bilinears~\cite{Nagel:2004sw,Maniatis_THDMStability},
\begin{equation}
K_0 = \varphi_1^{\dagger} \varphi_1 + \varphi_2^{\dagger} \varphi_2, \quad
\tvec{K} =
\begin{pmatrix}
K_1 \\ K_2 \\ K_3
\end{pmatrix}
=
\begin{pmatrix}
\varphi_1^\dagger \varphi_2 + \varphi_2^\dagger \varphi_1\\
i \varphi_2^\dagger \varphi_1 - i \varphi_1^\dagger \varphi_2\\
\varphi_1^{\dagger} \varphi_1 - \varphi_2^{\dagger} \varphi_2
\end{pmatrix}.
\end{equation}
In terms of these bilinears $K_0$, $\tvec{K}$, the 
Higgs potential~\eqref{V_fields} reads
\begin{equation}
V = \xi_0 K_0 + \tvec{\xi}^\trans \tvec{K}
  + \eta_{00} K_0^2
  + 2 K_0\tvec{\eta}^\trans \tvec{K}
  + \tvec{K}^\trans E \tvec{K}
	\label{eq:potential}
\end{equation}
with parameters~$\xi_0$, $\eta_{00}$,
3-component vectors $\tvec{\xi}$, $\tvec{\eta}$
and a $3 \times 3$ matrix $E=E^\trans$, all real.

The standard CP transformation of the Higgs-doublet fields is defined by
\begin{equation}
\label{eq-simCP}
\varphi_i(x) \rightarrow \varphi_i^*(x')\,, \qquad i=1,2\,\,
\qquad x'=(x^0, -\tvec{x}).
\end{equation}
In terms of the bilinears, this standard CP~transformation 
is~\cite{Nishi:2006tg,Maniatis_CPViolationGeometry}
\begin{equation}
\label{eq-simCPK}
K_0(x) \rightarrow K_0(x')\;,\quad
\tvec{K}(x) \rightarrow \bar{R}_2 \tvec{K}(x')
\end{equation}
where $\bar{R}_2=\diag(1,-1,1)$, corresponding in $K$~space
to a reflection 
on the 1--3 plane.
Generalised CP transformations~(GCPs) are \mbox{defined} by
\cite{Lee:1966ik,Ecker:1981wv,Ecker:1983hz}
\begin{equation}
\varphi_i(x) \rightarrow U_{ij} \; \varphi_j^*(x'),
\quad i,j=1,2\,,
\end{equation}
with $U$ being an arbitrary unitary $2 \times 2$ matrix.
In terms of the bilinears this reads~\cite{Maniatis_CPViolationGeometry}
\begin{equation}
K_0(x)  \rightarrow K_0(x'),\quad
\tvec{K}(x) \rightarrow \bar{R}\; \tvec{K}(x')
\label{eq:GCP}
\end{equation}
with an improper rotation matrix~$\bar{R}$.

Requiring $\bar{R}^2=\unitmatrix_3$ leads to two types
of GCPs. In $K$ space:
\begin{alignat}{2}
\label{eq-pointref}
&(i)\phantom{i} \quad \bar{R}=-\unitmatrix_3, \quad &&\text{point reflection,}\\
&(ii) \quad \bar{R}= R^\trans \; \bar{R}_2\; R,\quad &&
\text{reflection on a plane}~(R \in SO(3) ).
\label{eq-planeref}
\end{alignat}

For a review of the bilinear formalism and its relation to the conventional field approach as well as the generalised CP transformations we refer to \cite{Ferreira_THDMGeometry}. There, also an extensive list of references is given.

While the CP transformations of type $(ii)$ in~\eqref{eq-planeref} are 
equivalent to the standard CP transformation~\eqref{eq-simCP}, 
respectively~\eqref{eq-simCPK}, up to a basis change, the point reflection transformation
of type $(i)$ is quite different and turns out
to  have very interesting properties.
Motivated by this geometric picture of
generalised CP transformations,
the most general THDM invariant under the point reflection~($i$) has
been studied in~\cite{Maniatis_MCPMFamiliesMassHierarchy,Maniatis_MCPMPhenomenology,Maniatis_MCPMPhenomenologyRadiativeEffects,Maniatis_MCPMPhenomenologyAddendum}.
Invariance of the THDM potential~\eqref{eq:potential} under the GCP transformation~\eqref{eq:GCP} with $\bar{R} = - \unitmatrix_3$ from~\eqref{eq-pointref} clearly requires
$\tvec{\xi}=\tvec{\eta}=0$, leading to
\begin{equation}
V_{\text{MCPM}} =
  \xi_0\, K_0 
  + \eta_{00}\, K_0^2
  + \tvec{K}^\trans\, E\, \tvec{K}\,.
\label{eq-potential}
\end{equation}
Without loss of generality the $3 \times 3$ matrix $E$ can be chosen to be diagonal
\begin{equation}
E = \text{diag}(\mu_1, \mu_2, \mu_3)
\end{equation}
with the ordering 
\begin{equation}
	\mu_1 \geq \mu_2 \geq \mu_3.
	\label{eq:MuOrdering}
\end{equation}
The conditions to obtain a physically acceptable theory are spelled out in~\cite{Maniatis_MCPMFamiliesMassHierarchy,Maniatis_MCPMPhenomenology} and read
\begin{equation}
\begin{aligned}
	\eta_{00} &> 0\,, \\
	\mu_a + \eta_{00} &> 0, \quad \text{for } a = 1,2,3\,, \\
	\xi_0 &< 0\,, \\
	\mu_3 &< 0\,.
\end{aligned}
\end{equation}
The potential~\eqref{eq-potential} of this model is, besides the point reflection symmetry of type $(i)$, 
invariant under three GCPs
of type $(ii)$; see~\eqref{eq-pointref}, \eqref{eq-planeref}.
Requiring also the 
Yukawa couplings to respect these four GCPs, it was found in~\cite{Maniatis_MCPMFamiliesMassHierarchy}
that at least two fermion families are 
necessary in order to have non-vanishing 
fermion masses. That is, a reason for
family replication was given. For two fermion families only three options for the Yukawa-coupling structure were found:
\begin{enumerate}[(a)]
	\item one massive and one massless family,
	\item two mass-degenerate families,
	\item large flavour changing neutral currents.
\end{enumerate}
The maximally CP-symmetric model (MCPM) was constructed as follows in \cite{Maniatis_MCPMFamiliesMassHierarchy,Maniatis_MCPMPhenomenology}. The option (a) above was chosen for the Yukawa couplings with the third family $(t,b,\tau)$ as massive one, the second family $(c,s,\mu)$ as massless one. The first family $(u,d,e)$ was added uncoupled to the Higgs fields. In this way the highly symmetric MCPM gives a first approximation to the fundamental mass spectrum as observed in nature; see section 5 of~\cite{Maniatis_MCPMFamiliesMassHierarchy}. The Yukawa couplings of the MCPM read
\begin{align}
\label{eq11}
\mathscr{L}_{\mathrm{Yuk}}(x) = 
  -c^{(1)}_{l\,3} & \;\Bigg\{
    \bar{\tau}_{R}(x)\,\varphi_1^\dagger(x)
    \begin{pmatrix} \nu_{\tau\,L}(x) \\ \tau_{L}(x) \end{pmatrix}
\notag\\ &
    -\bar{\mu}_{R}(x)\,\varphi_2^\dagger(x)
    \begin{pmatrix} \nu_{\mu\,L}(x) \\ \mu_{L}(x) \end{pmatrix}
    \Bigg\}
\notag\\
  +c^{(1)}_{u\,3} &\;\Bigg\{
    \bar{t}_{R}(x)\,\varphi_1^\trans(x)\,\epsilon
    \begin{pmatrix} t_{L}(x) \\ b_{L}(x) \end{pmatrix}
\notag\\ &
    -\bar{c}_{R}(x)\,\varphi_2^\trans(x)\,\epsilon
    \begin{pmatrix} c_{L}(x) \\ s_{L}(x) \end{pmatrix}
    \Bigg\}
\notag\\
  -c^{(1)}_{d\,3} &\;\Bigg\{
    \bar{b}_{R}(x)\,\varphi_1^\dagger(x)
    \begin{pmatrix} t_{L}(x) \\ b_{L}(x) \end{pmatrix}
\notag\\ &
    -\bar{s}_{R}(x)\,\varphi_2^\dagger(x)
    \begin{pmatrix} c_{L}(x) \\ s_{L}(x) \end{pmatrix}
    \Bigg\}
		+ h.c.
\end{align}
where $c^{(1)}_{l\,3}$, $c^{(1)}_{u\,3}$ and $c^{(1)}_{d\,3}$ 
are real positive constants, determined by the fermion masses
as discussed below.

In the unitary gauge electroweak symmetry breaking (EWSB) gives
\begin{equation}
\varphi_1 (x) =
\frac{1}{\sqrt{2}}
\begin{pmatrix}
0 \\ v_0 + \rho' (x)
\end{pmatrix}\;,
\quad
\varphi_2 (x) =
\begin{pmatrix}
H^+ (x) \\
\frac{1}{\sqrt{2}} ( h'(x) + i h''(x) )
\end{pmatrix}
\end{equation}
with the standard vacuum expectation value
\begin{equation}
	v_0 = \sqrt{\frac {-\xi_0} {\eta_{00} + \mu_3}} \approx 246 \text{ GeV.}
	\label{eq:VacuumExpectationValue}
\end{equation}
The physical Higgs-boson fields are
$\rho'$, $h'$, $h''$, $H^+$, and $H^- = {\left(H^+ \right)}^\dagger$. Their masses -- at tree level -- read \cite{Maniatis_MCPMFamiliesMassHierarchy,Maniatis_MCPMPhenomenology}
\begin{equation}
\begin{split}
m_{\rho'}^2\;\;&= 2 (- \xi_0)\;,\\
m_{h'}^2\;\;&= 2 v_0^2 (\mu_1 - \mu_3)\;,\\
m_{h''}^2\;&= 2 v_0^2 (\mu_2 - \mu_3)\;,\\
m_{H^\pm}^2&= 2 v_0^2 (- \mu_3)\;
\end{split}
\label{eq21}
\end{equation}
where the ordering of the masses
\begin{equation}
\label{eq23}
m_{h'}^2 \geq m_{h''}^2\;
\end{equation}
is predicted by the theory; see~\eqref{eq:MuOrdering}. Requiring now that the neutral Higgs bosons $h'$ and $h''$ are neither massless nor mass degenerate leads to the condition
\begin{equation}
	\mu_1 > \mu_2 > \mu_3,
\end{equation}
replacing the weaker one in~\eqref{eq:MuOrdering}.
The masses of $\tau$, $t$, $b$ are given by
\begin{equation}
\begin{split}
m_\tau &= c^{(1)}_{l\,3} \frac{v_0}{\sqrt{2}}\;,\\
m_t &= c^{(1)}_{u\,3} \frac{v_0}{\sqrt{2}}\;,\\
m_b &= c^{(1)}_{d\,3} \frac{v_0}{\sqrt{2}}\;.
\end{split}
\label{eq21end}
\end{equation}
This fixes the constants of the Yukawa couplings in~\eqref{eq11} in terms of the third-family fermion masses.

The complete Lagrangian after EWSB is given in appendix A of \cite{Maniatis_MCPMPhenomenology} and is implemented as Monte Carlo program for the Feynman rules in appendix A of the present paper.

In total, there are 11 parameters in the MCPM Lagrangian. Before EWSB the natural parameters are the gauge couplings $g$, $g'$, $g_s$, the Higgs potential parameters $\mu_1$, $\mu_2$, $\mu_3$, $\xi_0$, $\eta_{00}$, and the Yukawa coefficients $c_{l\,3}^{(1)}$, $c_{u\,3}^{(1)}$, $c_{d\,3}^{(1)}$. It is convenient to replace the above 11 parameters by another set of 11 independent parameters containing more directly and well measurable quantities. These are the electromagnetic and strong coupling constants $\alpha$ and $\alpha_s$, respectively, the Fermi constant $G_F$, the $Z$-boson mass $m_Z$, the fermion masses $m_t$, $m_b$, $m_\tau$, and the Higgs boson masses $m_{\rho'}$, $m_{h'}$, $m_{h''}$, $m_{H^\pm}$. The relation of these 11 parameters to the original ones is given in appendix A. The masses of the physical Higgs bosons are constrained by~\eqref{eq23} and from the experimental results on the so called oblique parameters in electroweak precision tests; see \cite{Maniatis_MCPMParameterSpace}.

Let us summarize the essential
properties of the MCPM:
\begin{itemize}
\item There are 5 physical Higgs particles, two CP even ones $\rho'$, $h'$, 
one CP odd one $h''$, and a charged Higgs-boson pair~$H^\pm$.
\item The $\rho'$ boson's fermionic couplings are exclusively to the third $(\tau, t, b)$~family.
The $\rho'$ behaves similarly to the SM Higgs boson.
\item The Higgs-boson-fermion couplings of $h'$, $h''$, $H^\pm$ are exclusively to
the second $(\mu, c, s)$~family with
strengths proportional to the masses of the third generation fermions.
\item The first $(e, u, d)$~family is uncoupled to the Higgs bosons.
\end{itemize}
For further details
we refer to~\cite{Maniatis_MCPMFamiliesMassHierarchy,Maniatis_MCPMPhenomenology,Maniatis_MCPMPhenomenologyRadiativeEffects,Maniatis_MCPMPhenomenologyAddendum,Maniatis_MCPMParameterSpace}.



\section{Predictions for Hadron Colliders}

\begin{figure}[t]
	\begin{minipage}[b]{0.45\textwidth}
		\centering
		\includegraphics[bb=282 170 760 417,width=\textwidth, clip]{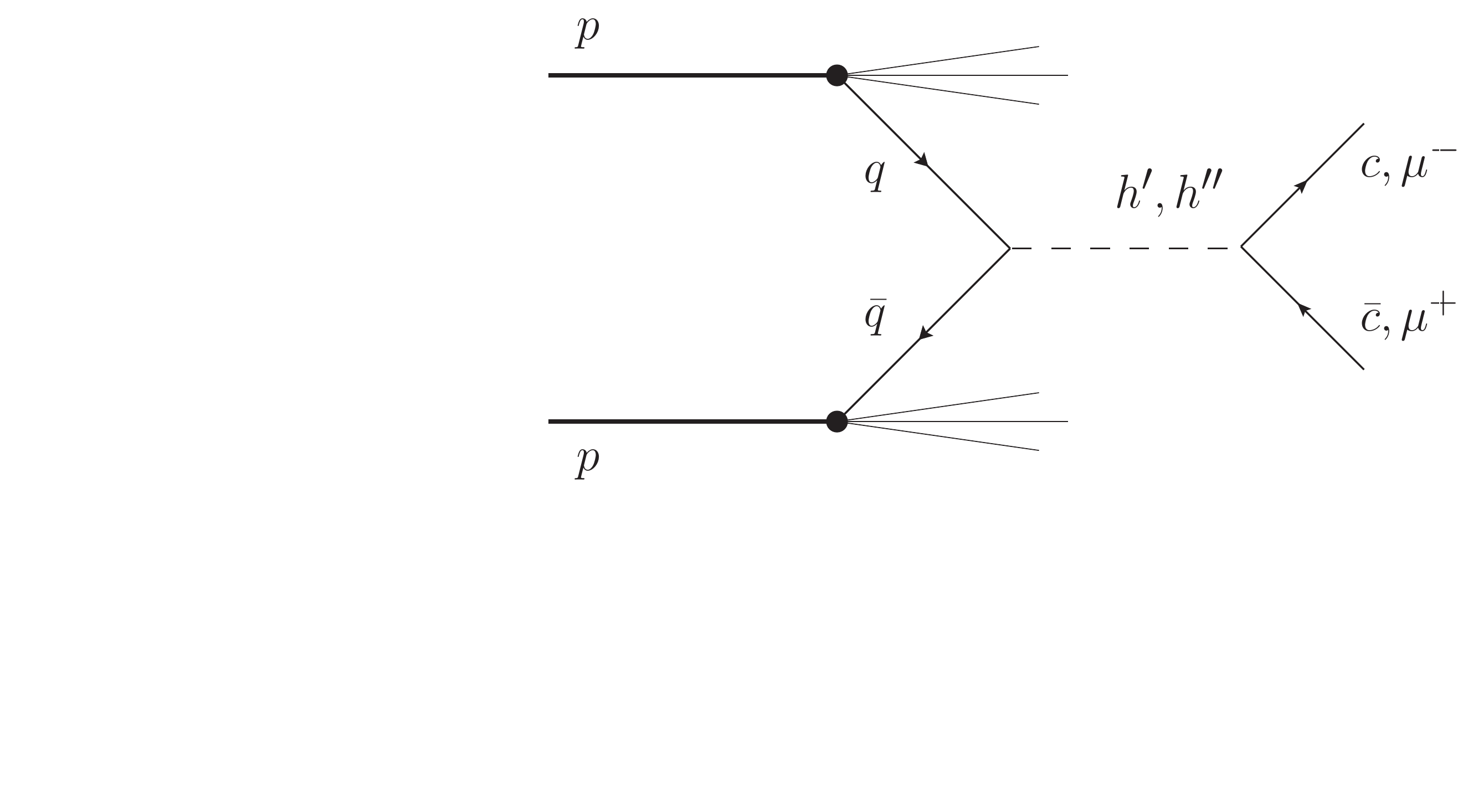}
	\end{minipage}
	\hfill
	\begin{minipage}[b]{0.45\textwidth}
		\centering
		\includegraphics[bb=282 170 760 417,width=\textwidth, clip]{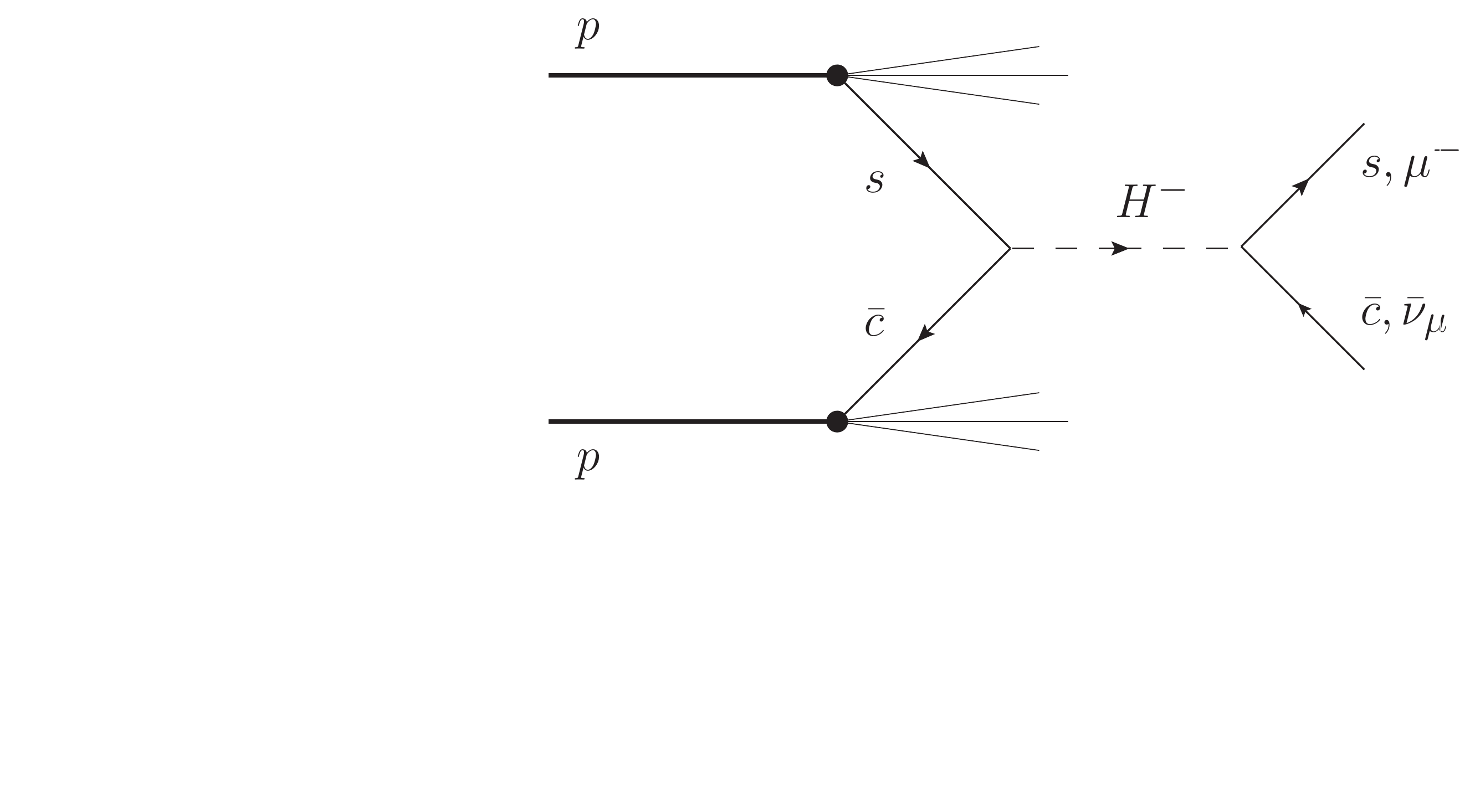}
	\end{minipage}
		\caption{Feynman diagrams for the dominant Drell--Yan type production processes for the Higgs bosons $h'$, $h''$ and $H^-$ in the MCPM ($q=c,s$). The dominant hadronic and leptonic decays are also indicated.}
		\label{fig-diag}
\end{figure}

\begin{figure}[t]
	\begin{minipage}[t]{0.45\textwidth}
		\centering
		\includegraphics[width=\textwidth]{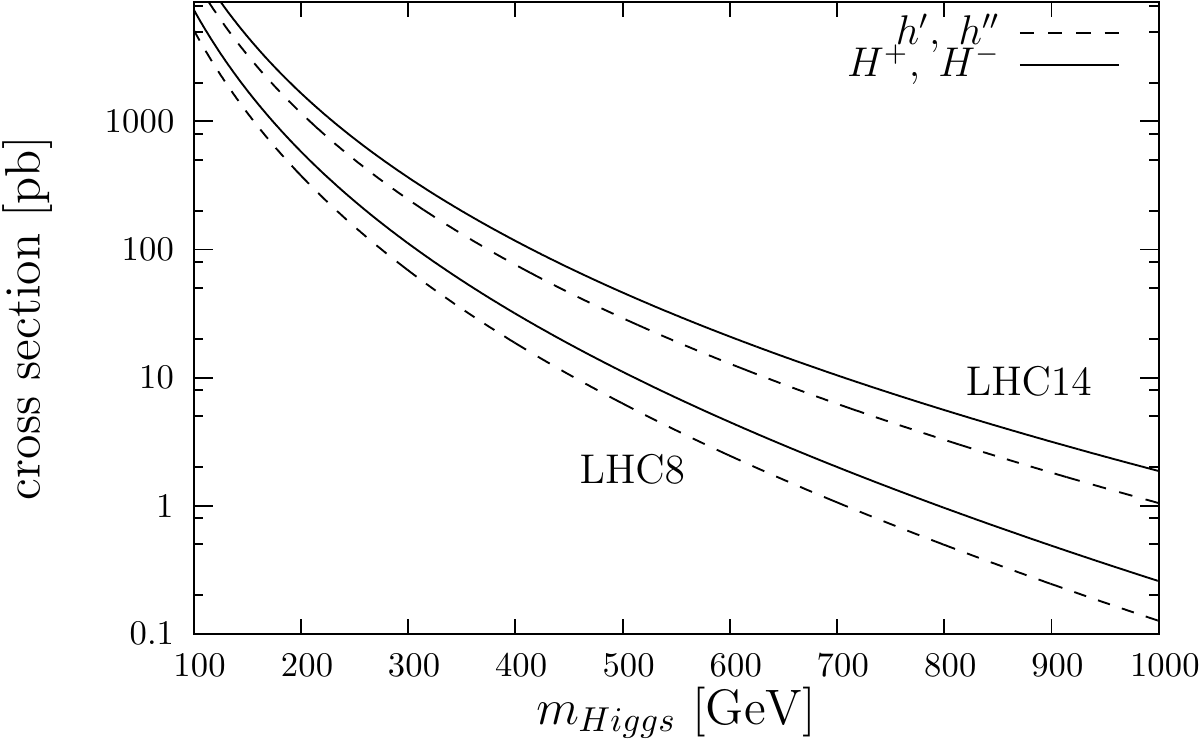}
		\caption{Total cross section of Drell--Yan type Higgs boson production at the LHC for c.\,m.~energies of 8~TeV and 14~TeV, respectively.}
		\label{fig-productioncrosssection}
	\end{minipage}
	\hfill
	\begin{minipage}[t]{0.45\textwidth}
		\centering
		\includegraphics[width=\textwidth]{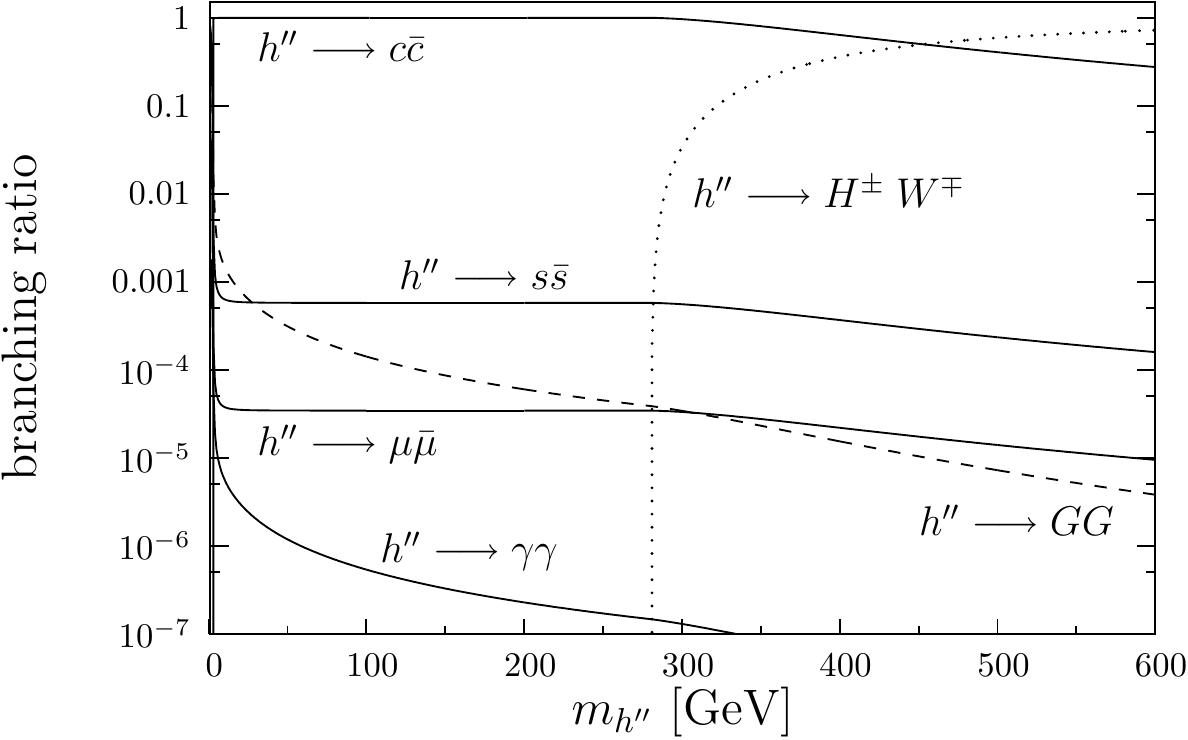}
		\caption{Branching ratios of the CP odd $h''$ Higgs boson, where a mass of $m_{H^\pm} = 200$ GeV is assumed. Figure taken from \citep{Maniatis_MCPMPhenomenology}.}
		\label{fig-branchingratios}
	\end{minipage}
\end{figure}

Since the Yukawa couplings of the
$h'$, $h''$, $H^\pm$ Higgs bosons to
the second fermion family are proportional
to the third-fermion-family masses we have for them large 
cross sections for Drell--Yan type production.
For the same reason we have large decay rates of
these Higgs bosons to the second generation fermions.
In figure~\ref{fig-diag} we show the diagrams
for these production and decay reactions for the $h'$, $h''$, and $H^-$ bosons in $pp$~collisions.
The corresponding diagram for $H^+$ production and decay is similar to the right one of figure~\ref{fig-diag} with the replacements $H^- \to H^+$, $s \to c$, $\bar c \to \bar s$, $\mu^- \to \nu_\mu$ and $\bar \nu_\mu \to \mu^+$.
In~\cite{Maniatis_MCPMPhenomenology,Maniatis_MCPMPhenomenologyAddendum}
the cross sections were computed for Drell--Yan Higgs-boson 
production at the LHC for center-of-mass
energies of 7~TeV and 14~TeV, respectively.
In~\cite{Maniatis_MCPMPhenomenologyRadiativeEffects} radiative effects were considered.
Here, we add the cross sections for a center-of-mass energy
of~8~TeV which is currently available at the LHC.
The corresponding total cross sections for the Drell--Yan 
production of the 
$h'$, $h''$, $H^\pm$ bosons
are shown in figure~\ref{fig-productioncrosssection}.
In figure~\ref{fig-branchingratios} we also recall the
branching ratios of the $h''$ boson decays.
In the following we shall develop all the necessary theoretical tools for detailed comparisons of the MCPM predictions with experimental data at the LHC. In particular, a Monte Carlo event generator for the MCPM shall be constructed.


\section{Implementation of Monte Carlo Event Generation}
\label{ch:implementation}

We have implemented the MCPM into the Monte Carlo event-generation package MadGraph 5\,/\,MadEvent and validated this implementation with different methods.

MadGraph 5 \citep{MadGraph5} is the latest version of the matrix-element generator MadGraph. It is bundled with the event-generation package MadEvent. For simplicity, both will be referred to as MadGraph in the following. Given an arbitrary process for any implemented model, the program lists all contributing tree-level diagrams. For each of these diagrams, events are randomly produced and the cross section or decay width is calculated. The output of such a simulation is a list of events in the Les Houches Event File (LHEF) standard \citep{LHEF}. For each event it includes the particles involved and their four-momenta. This can be read by programs such as PYTHIA \citep{PYTHIA}, with which hadronisation effects can be simulated. 


MadGraph is a flexible framework and allows the implementation of new models through the Universal FeynRules Output (UFO) format \citep{UFO}. Its major downside is the limitation to tree-level processes. However, the most interesting and relevant processes in the MCPM do not involve any loops \citep{Maniatis_MCPMPhenomenology}, so this drawback is acceptable for our present study.

The implementation includes all particles and couplings of the MCPM. This not only allows the analysis of signal processes involving Higgs bosons, but also the simulation of any tree-level background process. The quantities listed at the end of section 2, namely
\begin{equation}
	\alpha,\ G_F,\ \alpha_s,\ m_Z,\ m_t,\ m_b,\ m_\tau,\ m_{\rho'},\ m_{h'},\ m_{h''} \text{, and } m_{H^\pm}\,,
	\label{eq:MCParameters}
\end{equation}
are used as independent parameters.

In addition, more parameters can be used for the study of SM background processes. At tree level the MCPM predicts massless fermions for the first and second generation and a unit CKM matrix, $V_{CKM} = \unitmatrix_3$. Of course, this is not what we observe in nature, but it may represent a first approximation. Indeed, the ratios of the masses of first and second generation fermions to the corresponding third generation ones are small and $V_{CKM}$ is close to the unit matrix; see (125) and (126) of \citep{Maniatis_MCPMFamiliesMassHierarchy}. It may thus be that the GCPs of the MCPM play the role of \emph{approximate} symmetries. Thus, it should be sensible to analyse a model that combines the scalar sector of the MCPM with fermions of non-zero mass and a CKM matrix $V_{CKM} \neq \unitmatrix_3$, as done already in \citep{Maniatis_MCPMPhenomenology}.

For this reason our MadGraph implementation of the MCPM lets the user set fermion masses for the first two families as well. This affects phase space calculations, but the interactions remain unchanged. In a similar way the CKM matrix can be set to an arbitrary matrix, which changes SM processes, but not the Higgs boson vertices. With these parameters the user can decide whether to analyse the strict MCPM or a hybrid theory with fermion masses and a SM-like CKM matrix.

The MadGraph implementation of the MCPM was validated in two different ways. First, total cross sections and decay widths were checked for a number of different processes and parameter sets. These include the production of the MCPM Higgs bosons via quark-antiquark fusion as well as their fermionic decays. In all processes and parameter configurations good agreement between Monte Carlo results and theoretical expectation was found.

Second, angular dependencies and invariant mass distributions were analysed for different processes including the Drell-Yan-type production of Higgs bosons followed by their decay into fermion pairs. Again, the event shapes agree well with the expected distributions. For more details see \citep{Brehmer_BScThesis}. 


\section{Analysis of MCPM Signatures}
\label{ch:analysis}

It is now straightforward to ask whether the MCPM can be discovered or excluded at the LHC. A full answer requires a thorough analysis including the simulation of fragmentation and detector behaviour, which goes beyond the scope of this publication. The hadronic decay modes will thus not be considered here. A data analysis would neccessiate the tagging of charm jets which is experimentally challenging in the presence of huge QCD backgrounds. A study using a parton--level MC generator would not yield sensible results.

However, there is no hadronisation for leptons. Muons, which play an important role in the MCPM 
can be reconstructed relatively precisely in experiments. Hence, for the muonic decay channels the hard-process events generated by MadGraph are worth a look even without fragmentation and detector simulation. As a first application we use the new MadGraph implementation to compare the muonic MCPM signatures to the SM background.

Two such channels are analysed. The first one consists of the production of $h'$ or $h''$ via quark-antiquark fusion with subsequent decay into a $\mu^- \mu^+$ pair. The dominant SM background is given by the $\gamma^*$ and $Z$ Drell-Yan processes. As second channel we analyse the production of a $H^-$ boson decaying into a $\mu^- \bar \nu_\mu$ final state. Here the dominant background stems from the production of $W^-$ bosons. The corresponding Feynman diagrams can be found in figure \ref{fig-diag}. Additional backgrounds from e.\,g.\ top production have not been considered. Experimental studies show that they are below 15\,\% for most of the tested phase space \citep{Collaboration:2011dca, Chatrchyan:2012it, Aad:2011yg, Chatrchyan:2012qk}.
The production and decay of $H^+$ bosons is analogous to $H^-$ bosons and could be analysed in the same way.

Signal and background processes are simulated separately. Due to the different helicity structure of the lepton pairs from gauge-boson decays on the one hand and from Higgs-boson decays on the other hand there is, nelecting the lepton masses, no interference of signal and background. The couplings of $\mu^+\mu^-$ to $\gamma^*$ and $Z$ are chirality conserving. Thus, in the limit $m_\mu = 0$ only the helicity combinations $(h_+, h_-) = (\frac 1 2, - \frac 1 2)$ and $(-\frac 1 2, \frac 1 2)$ occur for the $\mu^+\mu^-$ pair. The $h'$ and $h''$ bosons, however, couple in a chirality-changing way to $\mu^+\mu^-$. Hence, the helicity combinations of $\mu^+\mu^-$ can only be $(h_+, h_-) = (\frac 1 2, \frac 1 2)$ and $(-\frac 1 2, -\frac 1 2)$. Similar considerations apply to $W^- \rightarrow \mu^- \bar \nu_\mu$ versus $H^- \rightarrow \mu^- \bar \nu_\mu$ and $W^+ \rightarrow \mu^+ \nu_\mu$ versus $H^+ \rightarrow \mu^+ \nu_\mu$.

Two different parameter sets for the Higgs-boson masses are used, see table~\ref{tbl:ParameterSets}. For all other particles, PDG recommendations \citep{PDG} are chosen. For each channel and parameter configuration, at least 200,000 events (1,000,000 events) are generated for the signal (background) processes. All simulations are performed for the LHC design energy $\sqrt{s} = 14$~TeV as well as its current center-of-mass energy $\sqrt{s} = 8$~TeV.

\begin{table}[tb]
	\centering
	\begin{tabular}{rrrrrrrrr}
	\toprule[0.12em]
	Parameter set & $m_{\rho'}$ & $m_{h'}$ & $\Gamma_{h'}$ & $m_{h''}$ & $\Gamma_{h''}$ & $m_{H^\pm}$ & $\Gamma_{H^\pm}$ \\
	\midrule
	A & 125 & 200 & 12.08 & 150 & 9.06 & 150 & 9.07\\
	B & 125 & 400 & 24.16 & 300 & 18.12 & 300 & 18.14\\
	\bottomrule[0.12em]
 	\end{tabular}
	\caption{Higgs-boson mass parameters used for the comparison of MCPM signatures with SM background. For $h'$, $h''$ and $H^\pm$ the widths as calculated from table~3 of \citep{Maniatis_MCPMPhenomenology} are also given. For the mass parameters chosen these widths are practically equal to the total widths. All numbers are in units of GeV.}
	\label{tbl:ParameterSets}
\end{table}

The analysis is first performed without applying any phase-space cuts. Of course, there is no detector that is able to detect and reconstruct every muon. For instance, in the ATLAS detector at CERN the muon trigger chambers only cover the pseudorapidity region $|\eta| < 2.4$; see 
\citep{ATLAS_DesignReport}. 
In addition, selection cuts on the transverse momentum of the muons are applied by the experiments in order to guarantee efficient triggering and background suppression; a typical cut value is $p_T > 25 \text{ GeV}$. Therefore, we  also perform the analysis requiring
\begin{subequations}
\label{eq:Cuts}
\begin{align}
	|\eta| &< 2.4, \\
	p_T &> 25 \text{ GeV}
\end{align}
for each muon. In the $\mu^- \bar \nu_\mu$ channel,
\begin{equation}
	E_T^\text{miss} > 25 \text{ GeV} 
\end{equation}
\end{subequations}
is also required, where $E_T^\text{miss}$ is the missing transverse energy due to the non--detection of the neutrino.

\begin{figure}[tbp]
	\centering
	\subfloat[]{
		\begin{footnotesize}
			\input{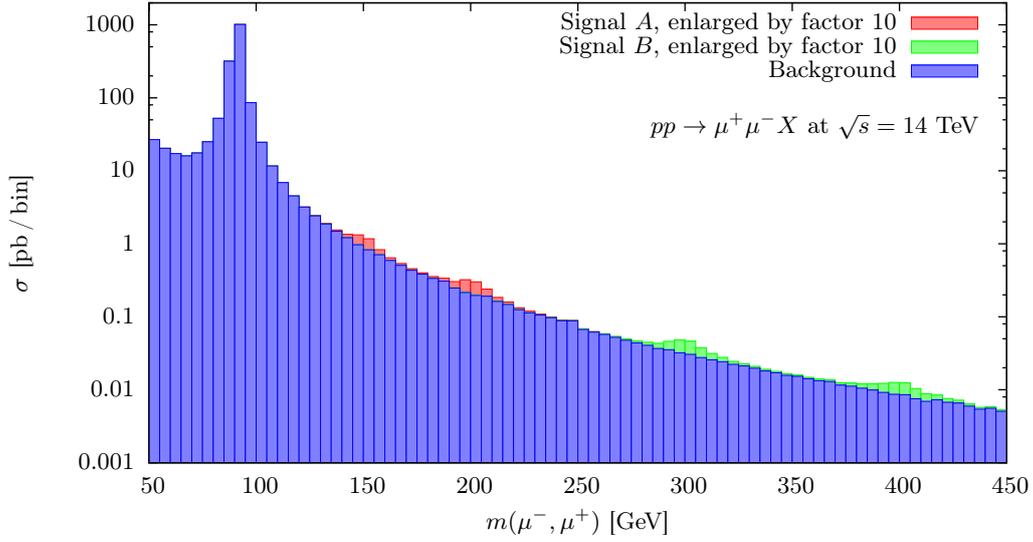}
			\label{fig:PlotMumu14Raw}
		\end{footnotesize}
	}\\
	\subfloat[]{
		\begin{footnotesize}
			\input{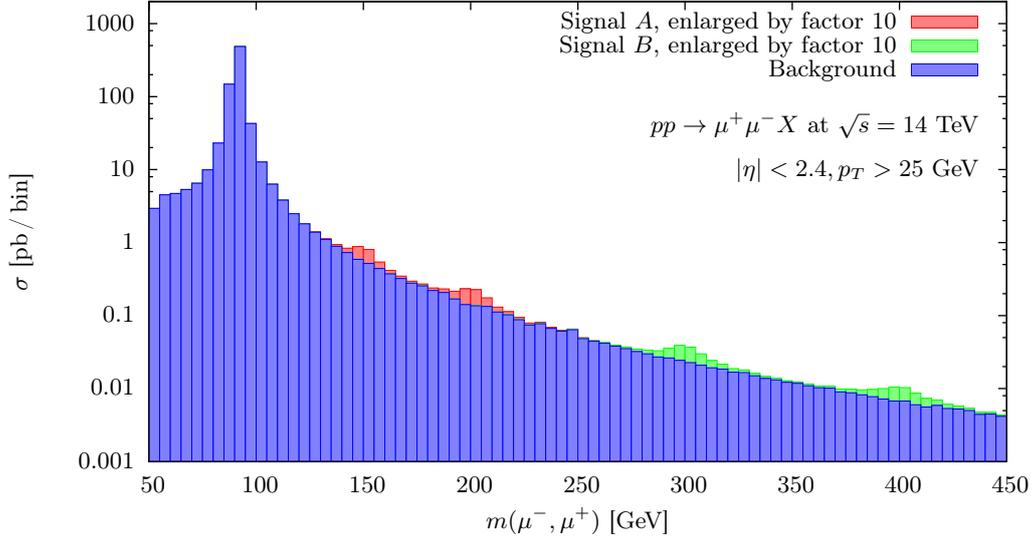}
			\label{fig:PlotMumu14Cuts}
		\end{footnotesize}
	}
	\caption{MCPM events compared to the SM background in the $\mu^+ \mu^-$ channel as a function of the invariant mass of the muon pair for $pp$ collisions at a center-of-mass energy of $\sqrt{s} = 14$~TeV. Two different parameter sets $A$ and $B$ are used for the MCPM, see table~\ref{tbl:ParameterSets}. (a) No selection cuts applied; (b) selection cuts on $\eta$ and $p_T$ of the muons applied.}
	\label{fig:PlotMumu14}
\end{figure}

\begin{figure}[tbp]
	\centering
	\subfloat[]{\begin{footnotesize}\input{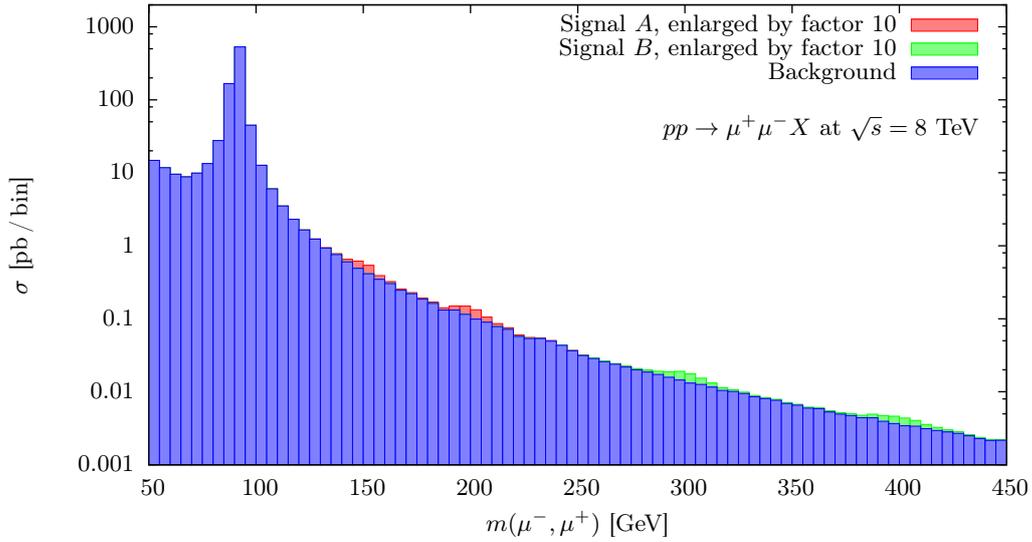} \label{fig:PlotMumu8Raw}\end{footnotesize}}\\
	\subfloat[]{\begin{footnotesize}\input{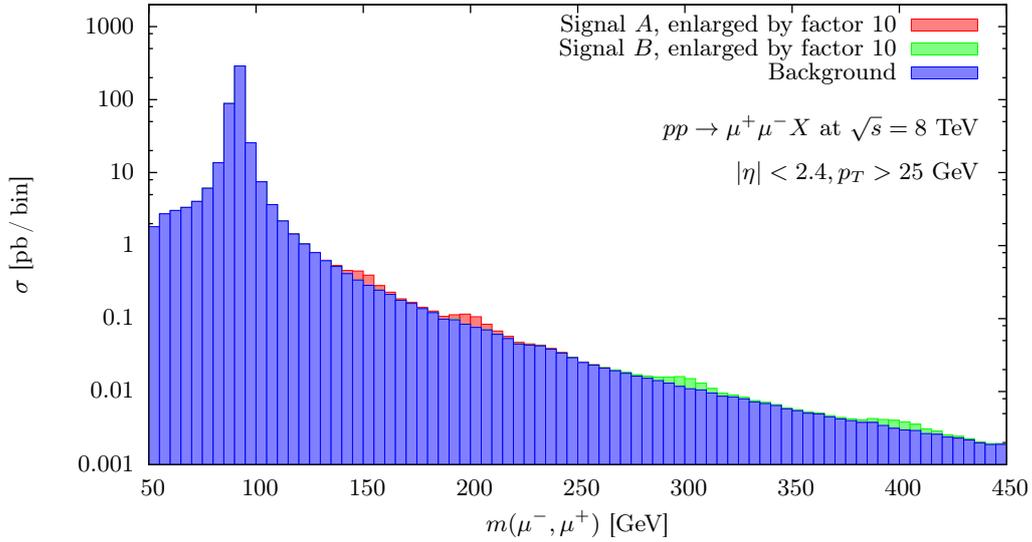} \label{fig:PlotMumu8Cuts}\end{footnotesize}}
	\caption{Same as figure~\ref{fig:PlotMumu14}, but with a center-of-mass energy of $\sqrt{s} = 8$~TeV. (a) No selection cuts applied; (b) selection cuts on $\eta$ and $p_T$ of the muons applied.}
	\label{fig:PlotMumu8}
\end{figure}

For the $\mu^- \mu^+$ channel, the resulting distribution of the invariant mass $m(\mu^-,\mu^+)$ of the muon pair is shown in figure~\ref{fig:PlotMumu14} for a center-of-mass energy of $\sqrt{s} = 14$~TeV and in figure~\ref{fig:PlotMumu8} for $\sqrt{s} = 8$~TeV. Here and in the following the bin size in mass is chosen as $\Delta m = 5$ GeV. Note that the resonance peaks from the Higgs bosons are enhanced by a factor 10. Thus, these Higgs-boson resonances are tiny in comparison to the background. The application of selection cuts and a higher center-of-mass energy improve the situation slightly. 

\begin{figure}[tbp]
	\centering
	\subfloat[]{\begin{footnotesize}\input{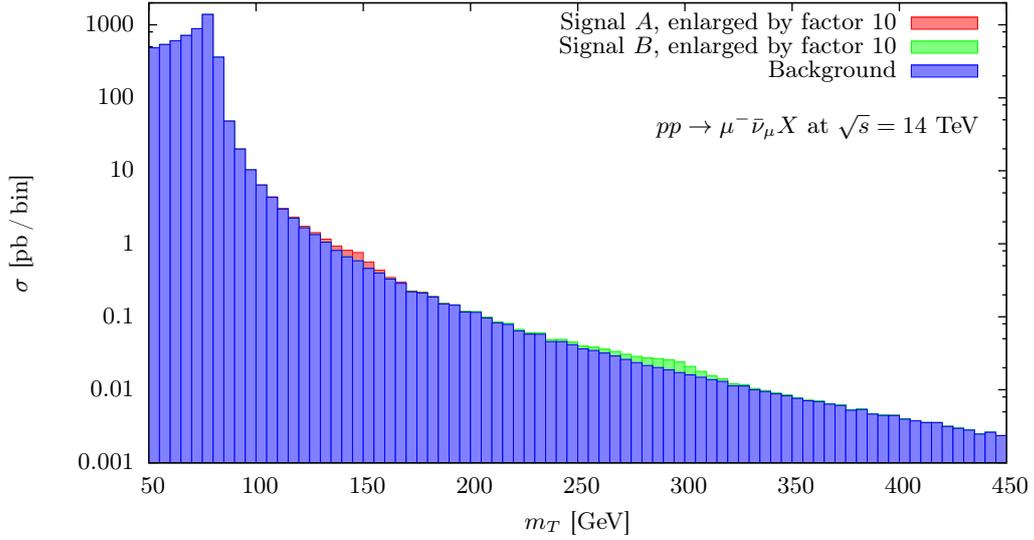} \label{fig:PlotMuv14Raw}\end{footnotesize}}\\
	\subfloat[]{\begin{footnotesize}\input{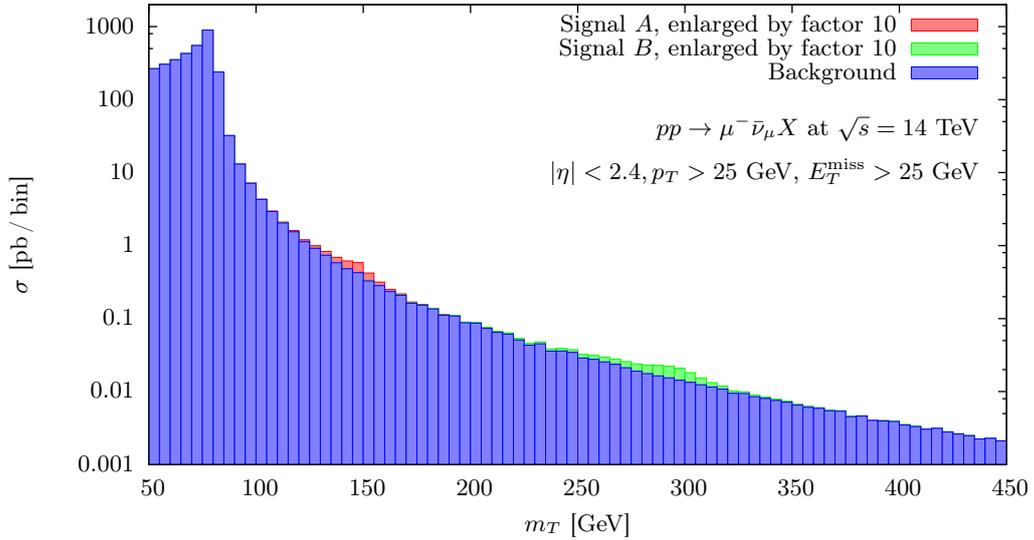} \label{fig:PlotMuv14Cuts}\end{footnotesize}}
	\caption{MCPM events compared to the SM background in the $\mu^- \bar \nu_\mu$ channel as a function of the transverse mass of the final state lepton pair for $pp$ collisions at a center-of-mass energy of $\sqrt{s} = 14$~TeV. Again the parameter sets $A$ and $B$ are used for the MCPM, see table~\ref{tbl:ParameterSets}. (a) No selection cuts applied; (b) selection cuts on $\eta$ and $p_T$ of the muon as well as $E^\text{miss}_T$ applied.}
	\label{fig:PlotMuv14}
\end{figure}

\begin{figure}[tbp]
	\centering
	\subfloat[]{\begin{footnotesize}\input{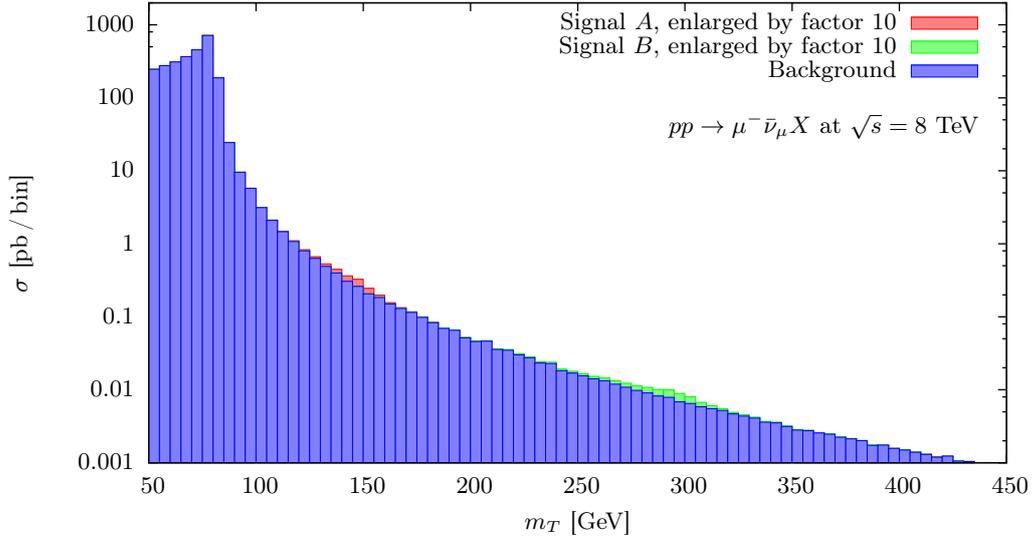} \label{fig:PlotMuv8Raw}\end{footnotesize}}\\
	\subfloat[]{\begin{footnotesize}\input{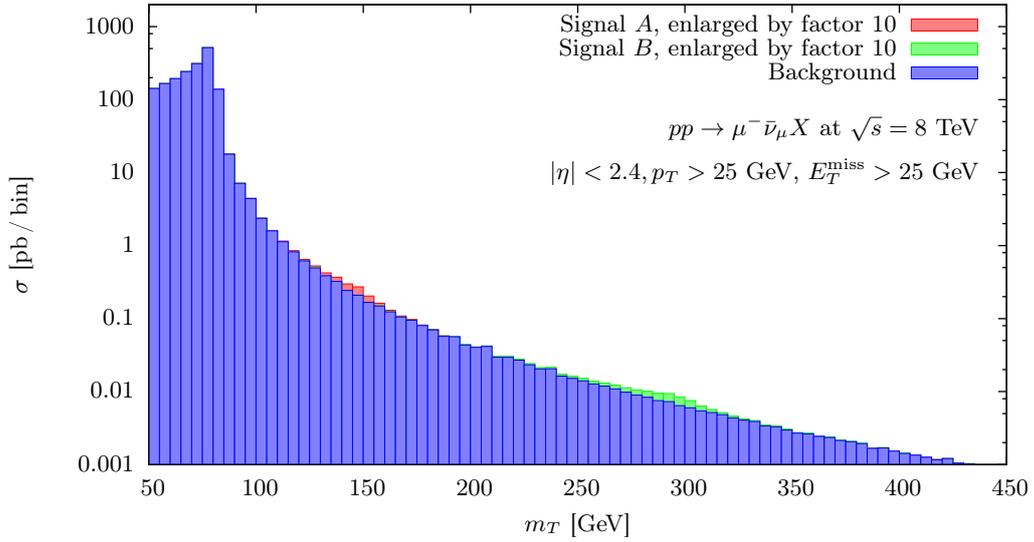} \label{fig:PlotMuv8Cuts}\end{footnotesize}}
	\caption{Same as figure~\ref{fig:PlotMuv14}, but with a center-of-mass energy of $\sqrt{s} = 8$~TeV. (a) No selection cuts applied; (b) selection cuts on $\eta$ and $p_T$ of the muon as well as $E^\text{miss}_T$ applied.}
	\label{fig:PlotMuv8}
\end{figure}

For the $\mu^- \bar \nu_\mu$ channel, the event distribution in the transverse mass $m_T$ of the lepton pair is given in figures~\ref{fig:PlotMuv14} and~\ref{fig:PlotMuv8} for $\sqrt{s} = 14$~TeV and 8~TeV, respectively. Again, the Higgs-boson resonances are much smaller than the background, even using appropriate selection cuts and a center-of-mass energy of 14~TeV.

Using the distributions after the application of selection cuts we calculate how much statistics is needed so that the MCPM resonance peaks are \emph{locally} significant compared to the background, i.\,e.\ larger than $n \sigma$, where $\sigma$ is the statistical uncertainty of the background. Conventionally, an excess with $n = 5$ is considered a discovery, while $n=2$ is often used for exclusion limits.

The detector resolution is a limiting factor in the search for resonances. Depending on the channel, the search strategy, the energy scale, the pseudorapidity region and a number of other parameters, the ATLAS collaboration quotes e.g. a design value for the dimuon invariant-mass resolution between $2$ and $12$ GeV; for the transverse-mass resolution the quoted values are slightly worse \citep{ATLAS_DesignReport}. Based on these figures and on the widths $\Gamma$ of the Higgs bosons as given in table~\ref{tbl:ParameterSets},
two to five bins were taken into account when calculating the integrated luminosities needed for a local 2$\sigma$-- and 5$\sigma$--significance, depending on the mass scale.

In table~\ref{tbl:LuminositiesNeeded} we give the integrated luminosities needed for local 2$\sigma$-- or 5$\sigma$--significance of the peaks due to the Higgs bosons $h'$, $h''$ and $H^-$ of the MCPM, as shown in figures~\ref{fig:PlotMumu14Cuts}, \ref{fig:PlotMumu8Cuts}, \ref{fig:PlotMuv14Cuts} and \ref{fig:PlotMuv8Cuts}. It should be stressed that these results are very rough estimates. On the one hand, only the local significance in one channel for the selection cuts given in~\eqref{eq:Cuts} is described. The sensitivity of the search might be improved by selection-cut optimisation for the different channels. On the other hand, only hard tree-level processes are taken into account and the behaviour of the detector was not simulated.

Keeping this caveat in mind, these numbers may nevertheless hint at the order of magnitude of statistics needed for MCPM signatures to become visible.
 


\begin{table}[tb]
	\centering
	\subfloat[$\sqrt{s} = 8$~TeV]{
		\begin{tabular*}{\textwidth}{@{\extracolsep{\fill}} llrrr}
			\toprule[0.12em]
			\multirow{2}{*}{Parameter set} & \multirow{2}{*}{Process} & \multirow{2}{*}{$m_H$ [GeV]} & \multicolumn{2}{c}{Necessary int.\ luminosity $\int L \text d t$ [$\text{fb}^{-1}$]}\\
			\cmidrule{4-5}
			& & & $2\sigma$ significance & $5\sigma$ significance\\
			\midrule
			A & $h'' \rightarrow \mu^+\mu^-$ & 150 & 5.2 & 33 \\
			A & $h' \rightarrow \mu^+\mu^-$ & 200 & 17 & 100 \\
			B & $h'' \rightarrow \mu^+ \mu^-$ & 300 & 99 & 620 \\
			B & $h' \rightarrow \mu^+ \mu^-$ & 400 & 490 & 3100 \\
			\midrule
			A & $H^- \rightarrow \mu^-\bar \nu_\mu$ & 150 & 9.3 & 58 \\
			B & $H^- \rightarrow \mu^-\bar \nu_\mu$ & 300 & 140 & 890 \\
			\bottomrule[0.12em]
		\end{tabular*}
		\label{tbl:Lum8TeV}
	} \\
	\subfloat[$\sqrt{s} = 14$~TeV]{
		\begin{tabular*}{\textwidth}{@{\extracolsep{\fill}} llrrr}
			\toprule[0.12em]
			\multirow{2}{*}{Parameter set} & \multirow{2}{*}{Process} & \multirow{2}{*}{$m_H$ [GeV]} & \multicolumn{2}{c}{Necessary int.\ luminosity $\int L \text d t$ [$\text{fb}^{-1}$]}\\
			\cmidrule{4-5}
			& & & $2\sigma$ significance & $5\sigma$ significance\\
			\midrule
			A & $h'' \rightarrow \mu^+\mu^-$ & 150 & 1.3 & 8.3 \\
			A & $h' \rightarrow \mu^+\mu^-$ & 200 & 3.4 & 21 \\
			B & $h'' \rightarrow \mu^+ \mu^-$ & 300 & 17 & 104 \\
			B & $h' \rightarrow \mu^+ \mu^-$ & 400 & 64 & 400 \\
			\midrule
			A & $H^- \rightarrow \mu^-\bar \nu_\mu$ & 150 & 3.1 & 20 \\
			B & $H^- \rightarrow \mu^-\bar \nu_\mu$ & 300 & 30 & 190 \\
			\bottomrule[0.12em]
		\end{tabular*}
		\label{tbl:Lum14TeV}
	}
	\caption{Integrated luminosities that are needed for local significances of the Higgs-boson resonances in the invariant mass distribution and transverse mass distribution of the muonic final states $\mu^+\mu^-$ and $\mu^- \bar \nu_\mu$, respectively. The figures for $H^+$ production and its decay to $\mu^+ \nu_\mu$ are similar to those for $H^-$.}
	\label{tbl:LuminositiesNeeded}
\end{table}



\section{Conclusions}
\label{ch:conclusions}

In this paper we presented the implementation of a two-Higgs-doublet model with maximal CP symmetry, the MCPM, into the Monte Carlo event-generation package MadGraph.

The MCPM is an extension of the Standard Model (SM) that is based on the requirement of invariance under certain generalised CP transformations. It features a fermion structure in which only the third-generation fermions are massive. Of course this does not describe nature precisely, but the model gives a first approximation of what has been observed. The theory predicts five physical Higgs bosons. Concerning their Yukawa couplings, four of these bosons couple only to the second-generation fermions, but with coupling strengths given by the third-generation fermion masses. At colliders such as the LHC, these four Higgs bosons will be produced mainly via quark-antiquark fusion. Their $\mu^+ \mu^-$, $\mu^+ \nu_\mu$ and $\mu^- \bar \nu_\mu$ decay modes seem most promising for a discovery. For a proper search for MCPM signatures a Monte Carlo simulation is needed.

We implemented the MCPM into MadGraph, allowing the calculation of cross sections and the random generation of events for arbitrary tree-level processes. The output is in the LHEF format and can be used for further analysis, for instance using PYTHIA and GEANT. Therefore, this MadGraph implementation can be used as the starting point for a full and thorough Monte Carlo simulation of the MCPM. It was validated successfully with different methods. The event generator is available from \citep{Brehmer_ModelFiles}.

The implementation was then used to compare the MCPM signatures to the SM background at LHC energies. This analysis was only done for the muonic channels and restricted to hard processes, so the results are only a rough approximation. 
For a Higgs boson $h''$ ($H^-$) of 300 GeV, an integrated luminosity of approximately 100~$\text{fb}^{-1}$ (140~$\text{fb}^{-1}$) at a center-of-mass energy $\sqrt{s} = 8$~TeV has to be collected so that a local $2\sigma$ significance becomes possible. On the other hand, for $h''$ ($H^-$) of mass 150~GeV the necessary integrated luminosities at $\sqrt{s} = 8$~TeV for local $2\sigma$ significance are only 5.2~$\text{fb}^{-1}$ (9.3~$\text{fb}^{-1}$). At present (August 2012), the ATLAS and CMS experiments have both collected data representing roughly 13~$\text{fb}^{-1}$ of integrated luminosity. This number is expected to rise up to 20~$\text{fb}^{-1}$ in the current data taking period. Thus, it seems that the exclusion of a part of the Higgs-boson mass range or the discovery of small local excesses hinting at the MCPM might be within reach in the near future.

Finally, we note that the recent announcement \citep{ATLAS_Higgs,CMS_Higgs} of a signal for a boson of mass 125~GeV is -- so far -- not only compatible with the SM Higgs boson, but also with the SM-like Higgs boson $\rho'$ of the MCPM. Only detailed comparisons of the decay channels of the discovered boson with theoretical predictions will allow to draw further conclusions. We also note that the bounds on the masses of the $h'$, $h''$ and $H^\pm$ bosons given in \citep{Maniatis_MCPMParameterSpace} depend on the mass of the $\rho'$. Identifying the new boson with the $\rho'$ of the MCPM, these bounds can be sharpened. Only figure 2 of \citep{Maniatis_MCPMParameterSpace} remains relevant where $m_{\rho'} = 125$ GeV was chosen. Our parameter sets A and B of table~\ref{tbl:ParameterSets} also correspond to this choice of the $\rho'$ mass.


\acknowledgments

Parts of chapters 1 and 2 of this paper rely on the article \citep{Maniatis_MCPMPhenomenologyAddendum} published in the proceedings of the conference "Physics at LHC 2010" held at DESY in 2010. The general permission to draw on the articles of these proceedings as stated there is acknowledged. We want to thank A.\ von Manteuffel for reading the manuscript and for useful discussions.

\bibliography{References}


\appendix
\section{Appendix}

\subsection{Feynman Rules of the MCPM}
\label{app:FeynmanRules}

In appendix A of \citep{Maniatis_MCPMPhenomenology} the MCPM Lagrangian after electroweak symmetry breaking is presented and the Feynman rules for several vertices are derived. The most important vertices and the corresponding expressions are given in table~\ref{tbl:FeynmanRules1} and table~\ref{tbl:FeynmanRules2} below. 

\begin{table}[p]
		\centering
		\begin{tabular*}{.75 \textwidth}{@{\extracolsep{\fill}} cc}
			\toprule[0.12em]
			Vertex & Corresponding expression \\
			\midrule
			
			\quad & \quad \\
			\quad \quad	\begin{fmffile}{Feynman1}
				\begin{fmfgraph*}(60,40)
					\fmfleft{i}
					\fmfright{o2,o1}
					\fmflabel{$t$}{i}
					\fmflabel{$\rho'$}{o1}
					\fmflabel{$t$}{o2}
					\fmf{fermion}{i,v}
					\fmf{dashes}{v,o1}
					\fmf{fermion}{v,o2}
				\end{fmfgraph*}
			\end{fmffile} \quad \quad
			& $\displaystyle -i \frac {m_t} {v_0}$ \\
			\quad & \quad \\
			
			\quad & \quad \\
			\quad \quad	\begin{fmffile}{Feynman2}
				\begin{fmfgraph*}(60,40)
					\fmfleft{i}
					\fmfright{o2,o1}
					\fmflabel{$b$}{i}
					\fmflabel{$\rho'$}{o1}
					\fmflabel{$b$}{o2}
					\fmf{fermion}{i,v}
					\fmf{dashes}{v,o1}
					\fmf{fermion}{v,o2}
				\end{fmfgraph*}
			\end{fmffile} \quad \quad
			& $\displaystyle -i \frac {m_b} {v_0}$ \\
			\quad & \quad \\
			
			\quad & \quad \\
			\quad \quad	\begin{fmffile}{Feynman3}
				\begin{fmfgraph*}(60,40)
					\fmfleft{i}
					\fmfright{o2,o1}
					\fmflabel{$\tau$}{i}
					\fmflabel{$\rho'$}{o1}
					\fmflabel{$\tau$}{o2}
					\fmf{fermion}{i,v}
					\fmf{dashes}{v,o1}
					\fmf{fermion}{v,o2}
				\end{fmfgraph*}
			\end{fmffile} \quad \quad
			& $\displaystyle -i \frac {m_\tau} {v_0}$ \\
			\quad & \quad \\
			
			\quad & \quad \\
			\quad \quad	\begin{fmffile}{Feynman4}
				\begin{fmfgraph*}(60,40)
					\fmfleft{i}
					\fmfright{o2,o1}
					\fmflabel{$c$}{i}
					\fmflabel{$h'$}{o1}
					\fmflabel{$c$}{o2}
					\fmf{fermion}{i,v}
					\fmf{dashes}{v,o1}
					\fmf{fermion}{v,o2}
				\end{fmfgraph*}
			\end{fmffile} \quad \quad
			& $\displaystyle i \frac {m_t} {v_0}$ \\
			\quad & \quad \\
			
			\quad & \quad \\
			\quad \quad	\begin{fmffile}{Feynman5}
				\begin{fmfgraph*}(60,40)
					\fmfleft{i}
					\fmfright{o2,o1}
					\fmflabel{$s$}{i}
					\fmflabel{$h'$}{o1}
					\fmflabel{$s$}{o2}
					\fmf{fermion}{i,v}
					\fmf{dashes}{v,o1}
					\fmf{fermion}{v,o2}
				\end{fmfgraph*}
			\end{fmffile} \quad \quad
			& $\displaystyle i \frac {m_b} {v_0}$ \\
			\quad & \quad \\
			
			\quad & \quad \\
			\quad \quad	\begin{fmffile}{Feynman6}
				\begin{fmfgraph*}(60,40)
					\fmfleft{i}
					\fmfright{o2,o1}
					\fmflabel{$\mu$}{i}
					\fmflabel{$h'$}{o1}
					\fmflabel{$\mu$}{o2}
					\fmf{fermion}{i,v}
					\fmf{dashes}{v,o1}
					\fmf{fermion}{v,o2}
				\end{fmfgraph*}
			\end{fmffile} \quad \quad
			& $\displaystyle i \frac {m_\tau} {v_0}$ \\
			\quad & \quad \\
			
			\bottomrule[0.12em]
		\end{tabular*}
	\caption{Feynman rules for the most relevant vertices in the MCPM, part 1.}
	\label{tbl:FeynmanRules1}
\end{table}

\begin{table}[p]
		\centering
		\begin{tabular*}{.75 \textwidth}{@{\extracolsep{\fill}} cc}
			\toprule[0.12em]
			Vertex & Corresponding expression \\
			\midrule
			
			\quad & \quad \\
			\quad \quad	\begin{fmffile}{Feynman7}
				\begin{fmfgraph*}(60,40)
					\fmfleft{i}
					\fmfright{o2,o1}
					\fmflabel{$c$}{i}
					\fmflabel{$h''$}{o1}
					\fmflabel{$c$}{o2}
					\fmf{fermion}{i,v}
					\fmf{dashes}{v,o1}
					\fmf{fermion}{v,o2}
				\end{fmfgraph*}
			\end{fmffile} \quad \quad
			& $\displaystyle \frac {m_t} {v_0} \gamma_5$ \\
			\quad & \quad \\
			
			\quad & \quad \\
			\quad \quad	\begin{fmffile}{Feynman8}
				\begin{fmfgraph*}(60,40)
					\fmfleft{i}
					\fmfright{o2,o1}
					\fmflabel{$s$}{i}
					\fmflabel{$h''$}{o1}
					\fmflabel{$s$}{o2}
					\fmf{fermion}{i,v}
					\fmf{dashes}{v,o1}
					\fmf{fermion}{v,o2}
				\end{fmfgraph*}
			\end{fmffile} \quad \quad
			& $\displaystyle - \frac {m_b} {v_0} \gamma_5$ \\
			\quad & \quad \\
			
			\quad & \quad \\
			\quad \quad	\begin{fmffile}{Feynman9}
				\begin{fmfgraph*}(60,40)
					\fmfleft{i}
					\fmfright{o2,o1}
					\fmflabel{$\mu$}{i}
					\fmflabel{$h''$}{o1}
					\fmflabel{$\mu$}{o2}
					\fmf{fermion}{i,v}
					\fmf{dashes}{v,o1}
					\fmf{fermion}{v,o2}
				\end{fmfgraph*}
			\end{fmffile} \quad \quad
			& $\displaystyle - \frac {m_\tau} {v_0} \gamma_5$ \\
			\quad & \quad \\
						
			\quad & \quad \\
			\quad \quad	\begin{fmffile}{Feynman10}
				\begin{fmfgraph*}(60,40)
					\fmfleft{i}
					\fmfright{o2,o1}
					\fmflabel{$s$}{i}
					\fmflabel{$H^-$}{o1}
					\fmflabel{$c$}{o2}
					\fmf{fermion}{i,v}
					\fmf{dashes_arrow}{v,o1}
					\fmf{fermion}{v,o2}
				\end{fmfgraph*}
			\end{fmffile} \quad \quad
			& $\displaystyle - i \frac {1} {\sqrt{2} v_0} \left[ m_t (1 - \gamma_5) - m_b (1 + \gamma_5) \right]$ \\
			\quad & \quad \\
			
			\quad & \quad \\
			\quad \quad	\begin{fmffile}{Feynman11}
				\begin{fmfgraph*}(60,40)
					\fmfleft{i}
					\fmfright{o2,o1}
					\fmflabel{$\mu^-$}{i}
					\fmflabel{$H^-$}{o1}
					\fmflabel{$\nu_\mu$}{o2}
					\fmf{fermion}{i,v}
					\fmf{dashes_arrow}{v,o1}
					\fmf{fermion}{v,o2}
				\end{fmfgraph*}
			\end{fmffile} \quad \quad
			& $\displaystyle i \frac {m_\tau} {\sqrt{2} v_0} (1 + \gamma_5)$ \\
			\quad & \quad \\
						
			\bottomrule[0.12em]
		\end{tabular*}
	\caption{Feynman rules for the most relevant vertices in the MCPM, part 2. There are many more, including couplings of gauge bosons to Higgs bosons and Higgs-boson self-couplings. The arrows on the $H^\pm$ lines denote the flow of negative charge.}
	\label{tbl:FeynmanRules2}
\end{table}

\subsection{Using the Monte Carlo Generator}
\label{app:ImplementationUsage}

The MCPM model in MadGraph is used in the same way as every MadGraph model, see \citep{MadGraph5} for an overview. First it has to be loaded. After starting the MadGraph binary in a shell this can be done with the command
\begin{verbatim}
	import model MCPM -modelname
\end{verbatim}
where the option \verb|-modelname| is needed to ensure the correct particle names. Then processes are defined and the event generation is started:
\begin{verbatim}
	generate p p > h1 > m- m+
	add process p p > h2 > m- m+
	output -f
	launch
\end{verbatim}
The "\verb|generate|" and "\verb|add process|" commands are used to specify the processes MadGraph has to evaluate, while "\verb|output|" and "\verb|launch|" start the event generation. The particle names used in the implementation are given in the following section. Before events are produced, MadGraph asks the user whether the standard parameters are to be used or if he wants to modify them.

During the simulation, an in-browser status page keeps the user informed. In the end, MadGraph gives out the contributing Feynman diagrams, the cross-sections or decay widths, the corresponding uncertainty and a compressed file containing the generated events in LHEF \citep{LHEF} format. This can be used for further analysis, for instance with MadAnalysis or PYTHIA \citep{PYTHIA}.

\subsection{MCPM Parameter Relations}

Here we give the relation -- at tree level -- between the original parameters of the MCPM Lagrangian and the parameters of~\eqref{eq:MCParameters} used in the event generator. With $v_0$ given in terms of the original parameters in~\eqref{eq:VacuumExpectationValue} we have
\begin{align}
	\alpha &= \frac 1 {4\pi} \frac {g^2 g'^2}{g^2+g'^2} \,, \\
	G_F &= \frac 1 {\sqrt 2} v_0^{-2} \,,\\
	\alpha_s &= \frac {g_s^2} {4\pi} \,,\\
	m_Z &= \frac 1 2 v_0 \sqrt{g^2+g'^2} \,,\\
	m_\tau &= c_{l\,3}^{(1)} \frac {v_0} {\sqrt 2} \,,\\
	m_t &= c_{u\,3}^{(1)} \frac {v_0} {\sqrt 2} \,,\\
	m_b &= c_{d\,3}^{(1)} \frac {v_0} {\sqrt 2} \,,\\
	m^2_{\rho'} &= 2 v_0^2 (\eta_{00} + \mu_3) \,,\\
	m^2_{h'} &= 2 v_0^2 (\mu_1 - \mu_3) \,,\\
	m^2_{h''} &= 2 v_0^2 (\mu_2 - \mu_3) \,,\\
	m^2_{H^\pm} &= 2 v_0^2 (- \mu_3) \,.
\end{align}

\subsection{List of Particles and Parameters in the MadGraph Implementation of the MCPM}
\label{app:ImplementationParticlesParameters}

A list of MCPM particles and parameters with their MadGraph names are presented in table~\ref{tbl:ImplementationParticles} and~\ref{tbl:ImplementationParameters}, respectively.

\begin{table}[!htb]
	\centering
		\begin{tabular*}{\textwidth}{@{\extracolsep{\fill}} ll}
			\toprule[0.12em]
			Particles & Names in MadGraph MCPM implementation \\
			\midrule
			Protons & \verb|p| \\
			\midrule
			Quarks & \verb|u d c s t b| \\
			Antiquarks & \verb|u~ d~ c~ s~ t~ b~| \\
			Leptons & \verb|e- m- tt- ve vm vt| \\
			Antileptons & \verb|e+ m+ tt+ ve~ vm~ vt~| \\
			\midrule
			Gauge bosons & \verb|a z w+ w- g| \\
			\midrule
			$\rho'$, $h'$, $h''$ & \verb|rho h1 h2| \\
			$H^\pm$ & \verb|h+ h-| \\
			\bottomrule[0.12em]
		\end{tabular*}
	\caption{The MCPM particles and their names in the MadGraph implementation.}
	\label{tbl:ImplementationParticles}
\end{table}

\begin{table}[!htb]
	\centering
		\begin{tabular*}{\textwidth}{@{\extracolsep{\fill}} lllrlr}
			\toprule[0.12em]
			Section & Parameter & MadGraph & Default & Unit & Strict MCPM\\
			\midrule
			Gauge & $\frac 1 \alpha$ & \verb|aEWM1| & 127.916 & &\\
			& $G_F$ & \verb|Gf| & 0.000011664 & $\text{GeV}^{-2}$ &\\
			& $\alpha_s$ & \verb|aS| & 0.1184 & &\\
			& $m_Z$ & \verb|MZ| & 91.1876 & GeV & \\
			\midrule
			CKM matrix & $\lambda_{WS}$ & \verb|lamWS| & 0.2253 & & 0\\
			& $A_{WS}$ & \verb|AWS| & 0.808 & &\\
			& $\rho_{WS}$ & \verb|rhoWS| & 0.132 & &\\
			& $\eta_{WS}$ & \verb|etaWS| & 0.341 & &\\
			\midrule
			Fermion masses & $m_c$ & \verb|MC| & 1.42 & GeV & 0 \\
			& $m_t$ & \verb|MT| & 172.9 & GeV & \\
			& $m_b$ & \verb|MB| & 4.67 & GeV & \\
			& $m_e$ & \verb|Me| & 0.00051100 & GeV & 0 \\
			& $m_\mu$ & \verb|MM| & 0.10566 & GeV & 0 \\
			& $m_\tau$ & \verb|MTA| & 1.7768 & GeV & \\
			\midrule
			Higgs-boson masses & $m_{\rho'}$ & \verb|mrho| & 125  & GeV & \\
			& $m_{h'}$ & \verb|mh1| & 200 & GeV & \\
			& $m_{h''}$ & \verb|mh2| & 150 & GeV & \\
			& $m_{H^\pm}$ & \verb|mhc| & 150 & GeV & \\			
			\bottomrule[0.12em]
		\end{tabular*}
	\caption{The MCPM parameters and their names in the MadGraph implementation. Note that some parameters are zero in the MCPM (see last column), however they can be set to other values in this model. This allows the correct simulation of SM background processes (see section 3). The CKM matrix is given in the Wolfenstein parametrisation~\citep{Wolfenstein}.}
	\label{tbl:ImplementationParameters}
\end{table}

The parameters may be a bit confusing, because the $c$, $e$, $\mu$ can be set to be massive, while the $u$, $d$ and $s$ are always massless. This does not have anything to do with the MCPM. As explained in section 3, some of the consequences of strict symmetry under generalised CP transformations have been dropped in this implementation, so it is generally possible to set masses for the first- and second-generation fermions. However, MadGraph treats the very light fermions as massless (even in the SM, where the theory certainly predicts something else). This does not change the physics at collider experiments in the TeV range, but it saves some computation time. Therefore the list of parameters includes the masses of those fermions of the first two generations where the masses are relevant for the calculations.

MadGraph also allows to set decay widths for all particles, which are named similarly to the mass parameters. This is important for further analysis with programs such as PYTHIA; these parameters do not influence the generation of events with MadGraph at all.

\end{document}